\documentclass[9pt,twocolumn,twoside]{pnas-new}

\newcommand{\beginsupplement}{%
 \setcounter{table}{0}
  \renewcommand{\thetable}{S\arabic{table}}%
  \setcounter{figure}{0}
  \renewcommand{\thefigure}{S\arabic{figure}}%
  \setcounter{equation}{0}
  \renewcommand{\theequation}{S\arabic{equation}}%
}

\templatetype{pnasresearcharticle} 

\title{Near-optimal protocols in complex nonequilibrium transformations}

\author[a,b,1]{Todd R. Gingrich}
\author[c]{Grant M. Rotskoff} 
\author[d,e]{Gavin E. Crooks}
\author[c,b,d,f]{Phillip L. Geissler}

\affil[a]{Physics of Living Systems Group, Department of Physics, Massachusetts Institute of Technology, 400 Technology Square, Cambridge, Massachusetts 02139}
\affil[b]{Department of Chemistry, University of California, Berkeley, California 94720}
\affil[c]{Biophysics Graduate Group, University of California, Berkeley, California 94720}
\affil[d]{Physical Biosciences Division, Lawrence Berkeley National Laboratory, Berkeley, California 94720}
\affil[e]{Kavli Energy NanoSciences Institute, Berkeley, California 94720}
\affil[f]{Chemical Sciences Division, Lawrence Berkeley National Laboratory, Berkeley, California 94720}

\leadauthor{Gingrich}

\significancestatement{Classical thermodynamics was developed to help design the best protocols for operating heat engines which remain close to equilibrium at all times.
Modern experimental techniques for manipulating microscopic and mesoscopic systems routinely access far-from-equilibrium states, demanding new theoretical tools to describe the optimal protocols in this more complicated regime.
Prior studies have sought, in simple models, the protocol which minimizes dissipation.
We use computational tools to investigate the diversity of low-dissipation protocols.
We show that optimal protocols can be accompanied by a vast set of near-optimal protocols, which still offer the substantive benefits of the optimal protocol.
While solving for the optimal protocol is typically difficult, computationally identifying a near-optimal protocol can be comparatively easy.}

\authorcontributions{T.R.G, G.M.R., G.E.C., and P.L.G. designed research, analyzed data, and wrote the paper.
T.R.G. and G.M.R. performed research.}
\authordeclaration{The authors declare no conflicts of interest.}
\correspondingauthor{\textsuperscript{1}To whom correspondence should be addressed. E-mail: toddging@mit.edu}


\renewcommand{\d}[2]{\frac{d #1}{d #2}} 
 
\newcommand{\avg}[1]{\left\langle #1 \right\rangle}

\newcommand{\dL}{\delta\Lambda}

\begin{abstract}
The development of sophisticated experimental means to control nanoscale systems has motivated efforts to design driving protocols which minimize the energy dissipated to the environment.
Computational models are a crucial tool in this practical challenge.
We describe a general method for sampling an ensemble of finite-time, nonequilibrium protocols biased towards a low average dissipation.  
We show that this scheme can be carried out very efficiently in several limiting cases.
As an application, we sample the ensemble of low-dissipation protocols that invert the magnetization of a 2D Ising model and explore how the diversity of the protocols varies in response to constraints on the average dissipation.
In this example, we find that there is a large set of protocols with average dissipation close to the optimal value, which we argue is a general phenomenon. 
\end{abstract}

\dates{This manuscript was compiled on \today}
\doi{\url{www.pnas.org/cgi/doi/10.1073/pnas.XXXXXXXXXX}}

\begin{document}

\verticaladjustment{-2pt}

\maketitle
\thispagestyle{firststyle}
\ifthenelse{\boolean{shortarticle}}{\ifthenelse{\boolean{singlecolumn}}{\abscontentformatted}{\abscontent}}{}

\dropcap{W}hen a system is guided gradually from one equilibrium state to another, the amount of heat dissipated into its surroundings is insensitive to the manner of driving.
In the more realistic case of an irreversible transformation in finite time, however, the dissipation can vary greatly from one driving protocol to another.
These basic tenets of thermodynamics have received renewed attention in recent years due to improved capabilities for manipulating systems at small scales~\cite{Smith1992, Liphardt2002, Collin2005, Martinez2016, Blickle2012, Toyabe2010, Jun2014} and advances in the theoretical understanding of nonequilibrium fluctuations~\cite{Jarzynski1997,Crooks1999,Spinney2013}.
In particular, many studies have sought to identify which finite-time protocols transform a system with the minimum amount of dissipation~\cite{Weinhold1975, Ruppeiner1979, Salamon1980, Schmiedl2007, Gomez2008, Then2008, Zulkowski2012, Zulkowski2015}.
Protocols which are optimal in this sense provide the most efficient route to measuring equilibrium free energy differences\textemdash in simulations and in experiments~\cite{Maragakis2008}\textemdash via the Jarzynski relation~\cite{Schmiedl2007, Dellago2013}.
More generally, low-dissipation protocols provide insight into the optimal design of nanoscale machines, both synthetic~\cite{Blickle2012, Martinez2016} and natural \cite{Oster2000}.

However, it remains challenging to identify the minimum-dissipation protocol for complex, many-body systems driven far from equilibrium, despite recent progress~\cite{Zhang2014,Rotskoff2015,Kappen2016}.
The difficulty of computing strictly optimal protocols motivates a pragmatic question: how large is the set of nearly optimal protocols?
In this paper we develop a framework to characterize that set. 
We introduce an entropy which indicates how many different protocols realize the same value of dissipation.
For low values of dissipation, this protocol entropy quantifies how prevalent the near-optimal protocols are, highlighting when the system may be efficiently driven in many different ways.
In analogy with common techniques of statistical physics, we present Monte Carlo methods to numerically compute the entropy by sampling protocols with a preference for low dissipation.
The samples generated by this procedure demonstrate the distinct ways in which the system can be driven while maintaining the expectation of low dissipation. 
Variation among the sampled protocols accentuates features that are unimportant for ensuring low dissipation; similarly, the lack of variation highlights features that are essential for this goal. 
These ideas and capabilities complement previous approaches to determining optimal control procedures, employing tools with many similar features~\cite{Kappen2005prl,Zhang2014,Kappen2016}.
We elaborate on the connections in the Discussion. 

\begin{figure}
\centering
\includegraphics[width=\linewidth]{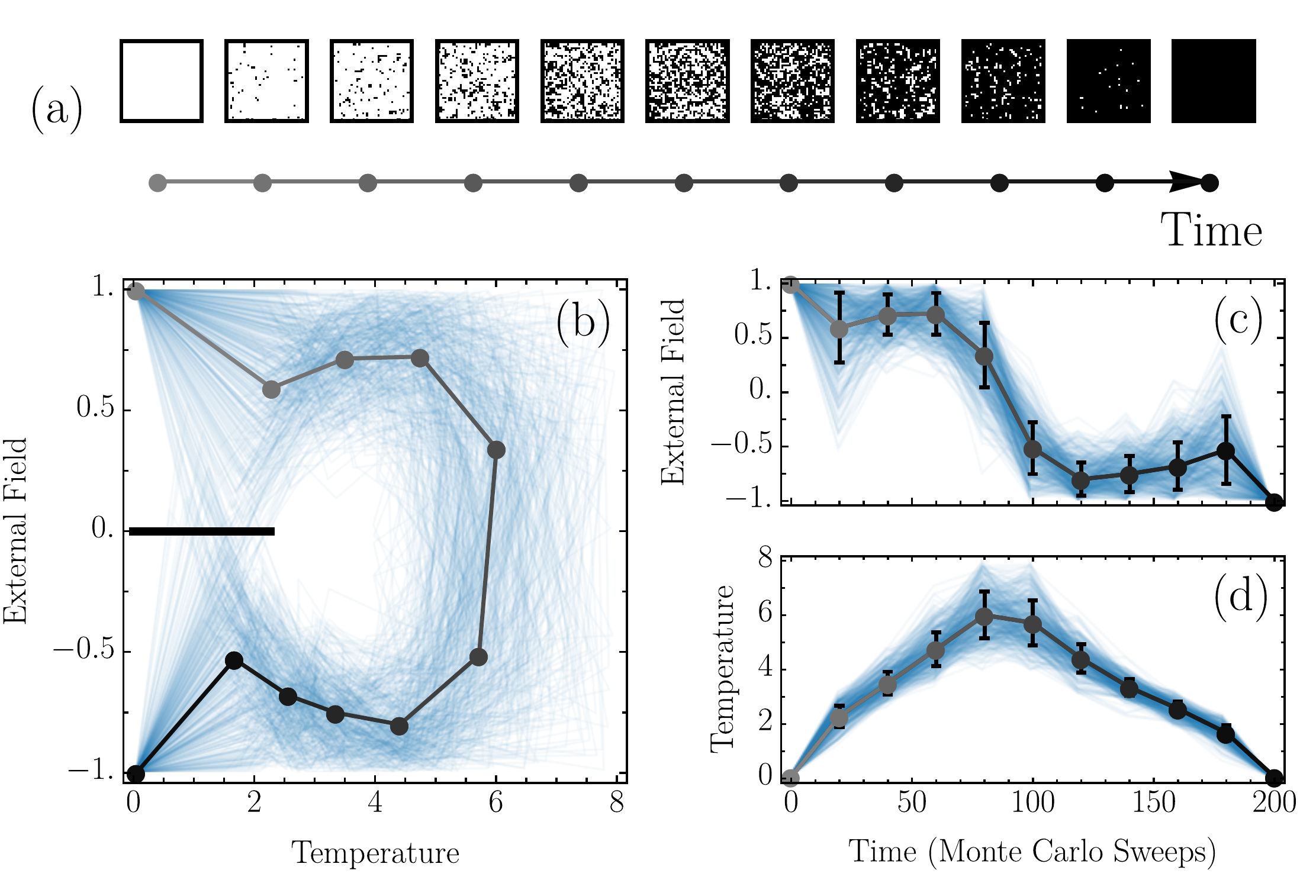}
\caption{\emph{Low-dissipation protocols that invert a 2D Ising magnet in finite time.}
(a) Snapshots of the $40\times 40$ periodically replicated magnet during field inversion.
(b) 450 representative samples of low-dissipation protocols (blue), collected from the $\lambda = 0.5, N = 5$ protocol ensemble,~\eqref{eq:Nprotocolbias}
Protocols are discretized into ten intervals in the temperature vs. field plane. 
Each interval represents a fixed amount of time (20 Monte Carlo sweeps) and the temperature and field values are linearly interpolated between the endpoints.
(c) The external field as a function of time.  (d) The temperature
as a function of time.  In panels (b)-(d) lines ranging from gray to
black indicate averages over the 450 protocols, with shading
corresponding to the times of the snapshots in (a).}
\label{fig:protocols}
\end{figure}

We illustrate our protocol-sampling framework with a numerical study of spin inversion of a ferromagnet, an essential process for copying information encoded in magnetic storage devices.
Reducing dissipation in this context is of practical interest because thermodynamic costs of copying and erasing bits is projected to account for a significant fraction of future computational energy expenditures~\cite{Lambson2011,Hong2016}.
We examine a simple microscopic model of this process, based on the two-dimensional Ising model (see Fig.~\ref{fig:protocols}). 
At low temperature and in the presence of an external magnetic field, spins align strongly in the direction of the field.
By adjusting the magnetic field and the temperature as functions of time, the magnetization may be rapidly inverted with a dissipation that depends on the manner in which the field and temperature are changed.
Analysis of the protocol entropy in this model indicates that a large set of non-optimal protocols can be used to control the system with a dissipation comparable to that of the optimal protocol.

\subsection*{Protocol entropy}

We first consider an ensemble of protocols $\Lambda(t)$ sharing the same value $\omega$ of average dissipation.
The protocol entropy $S(\omega)$ of this ensemble measures the density of protocols which have mean dissipation $\omega$.
In analogy to the standard microcanonical ensemble of statistical mechanics, we write the entropy as
\begin{equation}
S(\omega) = \ln \left[\Omega_0 \int \mathcal{D}\Lambda(t) \ \delta\left(\omega - \left<\omega\right>_\Lambda\right)\right],
\label{eq:protocolentropy}
\end{equation}
where the integral runs over the space of time-dependent protocols $\Lambda(t)$ and $\Omega_0$ is a constant which sets the arbitrary zero of entropy.
The delta function picks out protocols whose average dissipation $\left<\omega\right>_\Lambda$ lies within an infinitesimal interval around the specified value of $\omega$.
The $\Lambda$ subscript denotes an average taken over the probability distribution $P_\text{traj}[x(t)| \Lambda(t)]$ of stochastic trajectories evolving under the fixed protocol $\Lambda(t)$,
\begin{equation}
\left<\omega\right>_\Lambda = \int \mathcal{D} x(t) \ P_\text{traj}[x(t)| \Lambda(t)] \ \omega[x(t), \Lambda(t)].
\label{eq:meandissipation}
\end{equation}
For a single trajectory $x(t)$ the dissipation~$\omega$ can be cast in terms of an imbalance between forward and reversed dynamics~\cite{Spinney2013} with,
\begin{align}
\omega[x(t), \Lambda(t)] &= \ln \frac{P_{\rm traj}[x(t) | \Lambda(t)]}{P_{\rm traj}[\tilde{x}(t) | \tilde{\Lambda}(t)]}.
\label{eq:dissipation}
\end{align}
Tildes signify time-reversal, so the numerator and denominator are probabilities of forward and reverse trajectories, respectively.

Optimal protocols, which carry some minimal dissipation, represent a small fraction of the possible protocols.
Consequently, the protocol entropy evaluated at this minimal dissipation is low.
As the dissipation increases, the entropy increases, and the growth rate reflects how flexibly low-dissipation protocols may be constructed.
In particular, rapid entropic growth near the minimum dissipation suggests that targeting an exact optimal protocol is both challenging and gratuitous.
Many other protocols, in practice, will perform comparably to the optimum. 

The limiting behavior of $S(\omega)$ near the minimum dissipation $\omega^*$ can be calculated exactly, assuming only that $\avg{\omega}_\Lambda$ depends smoothly on variations in the low-dissipation protocols.
As shown in the Supporting Information (SI), protocol entropy grows logarithmically in this regime,
\begin{equation}
S(\omega) = \text{const} + \left( \frac{n}{2} - 1\right) \ln (\omega - \omega^*),
\label{eq:asymp}
\end{equation}
where $n$ is the total number of protocol degrees of freedom, i.e., the total number of parameter values defining any given protocol. 
Quite generally, therefore, protocol entropy increases sharply over a narrow range of dissipation values just above the minimum $\omega^*.$

\subsection*{A canonical protocol ensemble}
To numerically compute the protocol entropy, it is useful to introduce a canonical protocol ensemble
\begin{equation}
P_{\rm canon}[\Lambda(t)] \propto e^{-\gamma \left<\omega\right>_\Lambda}.
\label{eq:protocolbias}
\end{equation}
In correspondence with the canonical ensemble of statistical mechanics, $\avg{\omega}_\Lambda$ acts as an effective energy for each protocol and $\gamma$ plays the role of inverse temperature, tuning the mean dissipation, i.e., the average of $\omega$ over the distribution $P_{\rm canon}$.
Searching for a protocol with strictly minimum dissipation amounts to a zero ``temperature'' ($\gamma \to \infty$) quench while near-optimal protocols are identified by large values of $\gamma$.

When using a sufficiently large $\gamma$, the samples reveal representative low-dissipation protocols. 
By using different biasing strengths $\gamma$ in Eq.~\eqref{eq:protocolbias}, we can learn about the characteristics of protocols with distinct values of average dissipation.
In addition, the protocols sampled with various choices of $\gamma$ can be combined to calculate $S(\omega)$ over a broad range.

\subsubsection*{Sampling protocols and trajectories}
In principle, the ensemble defined by~\eqref{eq:protocolbias} may be directly sampled with a Monte Carlo procedure that conditionally accepts protocol changes based on the corresponding changes in $\avg{\omega}_{\Lambda}$. 
For complex systems, however, values of $\avg{\omega}_\Lambda$ are typically not known exactly.
They can be estimated from the sample mean~$\overline{\omega}_\Lambda = N^{-1} \sum_i \omega[x_i(t)|\Lambda(t)]$ of a collection of $N$ trajectories drawn from $P_\text{traj}[x(t)|\Lambda(t)]$.
But for finite $N$, replacing $\left<\omega\right>_\Lambda$ by $\overline{\omega}_\Lambda$ in~\eqref{eq:protocolbias} yields a distribution of protocols which differs from $P_\text{canon}[\Lambda(t)]$.
Strategies to correct for the finite-$N$ bias have been formulated to enable conventional Boltzmann sampling when configurational energies cannot
be calculated with certainty~\cite{Andrieu2009,Ball2003,Beaumont2003,Ceperley1999,Lin2000}. 
Here, we consider an analogous strategy in the context of sampling protocols. 

To sample the canonical protocol distribution, we construct a Monte Carlo procedure which performs a random walk through the joint space of protocols and $N$ independent trajectories $x_1(t), x_2(t), \hdots x_N(t)$.  
A trial move amounts to an attempt to make changes in both $\Lambda(t)$ and in $\left\{x_i(t)\right\}$.
Operationally, this proposed change can be achieved by first perturbing the protocol (with a symmetric generation probability) before regenerating the trajectories using the new protocol.
For simplicity, we consider the case that the trajectory generation probabilities are also symmetric, as in the noise-guided shooting procedures of transition path sampling~\cite{Dellago1998,Crooks2001,Hartmann2014,Gingrich2015}.
Accepting a trial move with the Metropolis probability $\min[1,\exp(-\lambda \Delta \overline{\omega})]$, where $\Delta \overline{\omega}$ is the difference between the sample means under the original and trial protocols, yields a stationary distribution
\begin{equation}
P_\text{sampled}[x_1(t), \dots, x_N(t), \Lambda(t)] \propto \exp \left( -\frac{\lambda}{N} \sum_{i=1}^N \omega[x_i(t)|\Lambda(t)]\right).
\label{eq:Nprotocolbias}
\end{equation}
The resulting marginal distribution of protocols,
\begin{equation}
P_\text{sampled}[\Lambda(t)] \propto \avg{ e^{-\lambda \omega / N} }_\Lambda^N,
\label{eq:marginal}
\end{equation}
is determined by the dissipation statistics of each protocol, but in a more complicated way than $P_\text{canon}$.
Nevertheless, we show that the sampled protocols are drawn from a canonical protocol distribution in two special situations: the case of Gaussian dissipation distributions and the large $N$ limit.

\subsubsection*{Cumulant expansion} 
The expression for the marginal distribution~\eqref{eq:marginal} may be recast as
\begin{equation}
P_\text{sampled}[\Lambda(t)]\propto e^{N \psi_\Lambda(-\lambda/N)},
\label{eq:psampledcumulant}
\end{equation}
where $\psi_\Lambda(k) = \ln \avg{e^{k\omega}}_\Lambda$ is the cumulant generating function for the dissipation.
A cumulant expansion,
\begin{equation}
\psi_\Lambda(-\lambda/N) = -\frac{\lambda \left<\omega\right>_\Lambda}{N} + \frac{\lambda^2 \left<\delta \omega^2\right>_\Lambda}{2N^2} - \mathcal{O}\left(\frac{\lambda^3}{N^3}\right),
\label{eq:cumulantexp}
\end{equation}
relates the probability of sampled protocols to the cumulants of the dissipation distribution $P(\omega|\Lambda(t))$, where $\delta\omega = \omega - \left<\omega\right>_\Lambda$.
As $N$ grows, the contribution from higher order cumulants vanishes, consistent with the central limit theorem. 
In the limit $N\to \infty$, sampling~\eqref{eq:protocolbias} becomes equivalent to sampling~\eqref{eq:Nprotocolbias} because the sample mean converges to the average dissipation.

In the special case that $P(\omega|\Lambda(t))$ is Gaussian for each protocol $\Lambda(t)$, a powerful simplification arises, averting the need to use large values of $N$.
Gaussian dissipation distributions occur in many contexts\textemdash as a defining feature of linear response~\cite{Speck2004}, in the limit of slow adiabatic driving~\cite{Speck2004}, and when Brownian particles evolve in driven harmonic potentials~\cite{Mazonka1999}.
In all these cases, the cumulants of $P(\omega|\Lambda(t))$ beyond the variance vanish, allowing us to exactly truncate~\eqref{eq:cumulantexp} at second order. 
If we further take $\Lambda(t)$ to be symmetric under time reversal, then the fluctuation theorem provides an exact relationship between the mean and variance:~$\left<\delta \omega^2\right>_\Lambda = 2 \left<\omega\right>_\Lambda$.
As a result, the biased protocol distribution can be expressed in terms of mean dissipation alone, 
\begin{equation}
P_\text{Gaussian}[\Lambda(t)] \propto e^{-\lambda\left(1-\lambda/N\right)\langle \omega\rangle_\Lambda}.
\label{eq:gaussianprotocol}
\end{equation}
\eqref{eq:gaussianprotocol} has precisely the form of the canonical protocol distribution~\eqref{eq:protocolbias}, with an effective bias $\gamma = \lambda(1-\lambda/N)$. 
This result offers tremendous flexibility.
An exact bias towards low average dissipation can be achieved with any $N$, e.g., by sampling a small number of trajectories for each proposed change in protocol. 
Since generating trajectories dominates the computational expense of our sampling scheme, the freedom to choose small $N$ is very attractive.

The limitation with using small $N$ when sampling protocols is that the achievable bias strength $\gamma$ cannot exceed $\gamma_{\rm max} = N/4$. 
This constraint arises because the $\lambda$ bias in~\eqref{eq:protocolbias} directly favors low-dissipation {\em trajectories}, and not necessarily low-dissipation {\em protocols}. 
For small values of $\lambda$, the trajectories sampled for a given protocol are typical of the unbiased trajectory distribution $P_{\rm traj}[x(t)|\Lambda(t)]$. 
In this case there is thus a strong correlation between sampled low-dissipation trajectories and protocols that yield low dissipation on average. 
This correspondence degrades for large $\lambda$. 
In fact, for $\lambda > N/2$, sampled trajectories have negative dissipation on average\footnote{The cumulant generating function $\psi_\Lambda(k)$ is symmetric about $k=-1/2$~\cite{Lebowitz1999}, and $P_{\rm sample}[\Lambda(t)]$ is correspondingly symmetric in $\lambda$ about $\lambda=N/2$. This symmetry implies negative average dissipation when $\lambda > N/2$.}, which cannot be typical of any protocol according to the second law of thermodynamics.
Thus, as $\lambda$ is increased toward $N/2$, the joint ensemble of trajectories and protocols switches from highlighting low-dissipation protocols to emphasizing rare negative-dissipation trajectories.  
Moreover, sampling with large values of $\lambda$ requires generation of increasingly rare trajectories, complicating efficient path sampling as discussed in the SI.

\section*{Results}
\subsection*{Spin inversion protocols}
To illustrate the use of the low-dissipation protocol ensemble, we consider the inversion of spins in a ferromagnet.
Specifically, we imagine initializing a system of interacting spins at low temperature, where its equilibrium state has long-range ``up'' or ``down'' order.
We then ask how best to flip this ``bit''.
That is, how should we vary the temperature and external field as functions of time to flip the state of the magnet without excess dissipation?
This problem, relevant to the design of low power magnetic hard drives, has been investigated as an optimal control problem elsewhere~\cite{Venturoli2009,Rotskoff2015}.
Here we also consider the near-optimal drivings.

We represent the ferromagnet as a two-dimensional Ising model with periodic boundary conditions and dynamics generated by a succession of individual spin flips.
With an external magnetic field $h$, the energy of a configuration is given by 
\begin{equation}
E = -h \sum_i \sigma_i - \sum_{\left<ij\right>} \sigma_i \sigma_j,
\label{eq:hamiltonian}
\end{equation}
where $\sigma_i = \pm 1$, and $\left<ij\right>$ indicates a sum over nearest neighbor sites $i$ and $j$.
An attempted spin flip that alters the energy by $\Delta E$ is accepted with Glauber probability $e^{-\Delta E/T}/(1 + e^{-\Delta E/T})$, where $T$ is the temperature of the bath. 
Unlike equilibrium Ising model dynamics, the temperature and magnetic field are time-dependent as prescribed by a nonequilibrium protocol~$\Lambda(t) = \left\{T(t), h(t)\right\}$.
In a finite amount of time $t_{\rm obs}$ we aim to switch from the macroscopic up state to the down state.
We therefore consider only protocols that begin and end at low temperature $[T(0) = T(t_{\rm obs}) = 0.05]$ and that switch from a positive to a negative field $[h(0) = -h(t_{\rm obs}) = 1]$.

One consequence of the nonequilibrium driving is that the dynamics is not microscopically reversible.
For ordinary Glauber dynamics, the equilibrium probability of a trajectory segment is equal to its time-reversed counterpart, but our time-dependent driving breaks this equality.
By tracking the random numbers which generate each spin flip, we explicitly compute the forward and time-reversed probabilities of each trajectory, thereby computing the stochastic thermodynamic dissipation via~\eqref{eq:dissipation}.
Physically, the dissipation of each microscopic step multiplied by $T$ is the heat transferred from the thermal bath into the system.

We use Monte Carlo techniques, discussed further in the Methods section, to sample low-dissipation protocols from~\eqref{eq:Nprotocolbias} with $\lambda = 0.5, N = 5$.
Fig.~\ref{fig:protocols} shows 450 representative protocols, all of which avoid the region of parameter space near the Ising critical point.
Control is particularly costly~\cite{Rotskoff2015} in this vicinity due to critical slowing down, which causes the spin system to lag behind changes in the control parameters.
There is a natural connection between dissipation and lag~\cite{Kawai2007,Vaikuntanathan2009}: the farther the system falls out of equilibrium with control parameters' instantaneous values, the more heat is dissipated to the reservoir during the relaxation.

Roughly, the optimal strategy requires that we first heat the magnet, next invert the field, and then cool the magnet. 
But the varied protocols in Fig.~\ref{fig:protocols} demonstrate significant leeway in how these steps are carried out.  
Most notably, while the system is held at low temperature, the magnetic field need not be precisely tuned, as evidenced by large variations both early and late in the protocol.  
Some low-dissipation protocols even transiently invert the field at low temperature, thereby crossing the equilibrium coexistence curve, only to restore the field's original sign a short time later. 
This seemingly wasteful procedure in fact incurs little dissipative cost, because it is highly ineffectual. 
The low-temperature field inversion is too brief for nucleation of the new phase to occur with significant probability, so the extent of relaxation is negligible. 
Absent relaxation, no heat is dissipated to the bath. 
Perhaps counterintuitively, the lag is so severe in this case as to be irrelevant.

\subsubsection*{Protocol Entropy} 
\begin{figure}[t]
\centering
\includegraphics[width=0.85\linewidth]{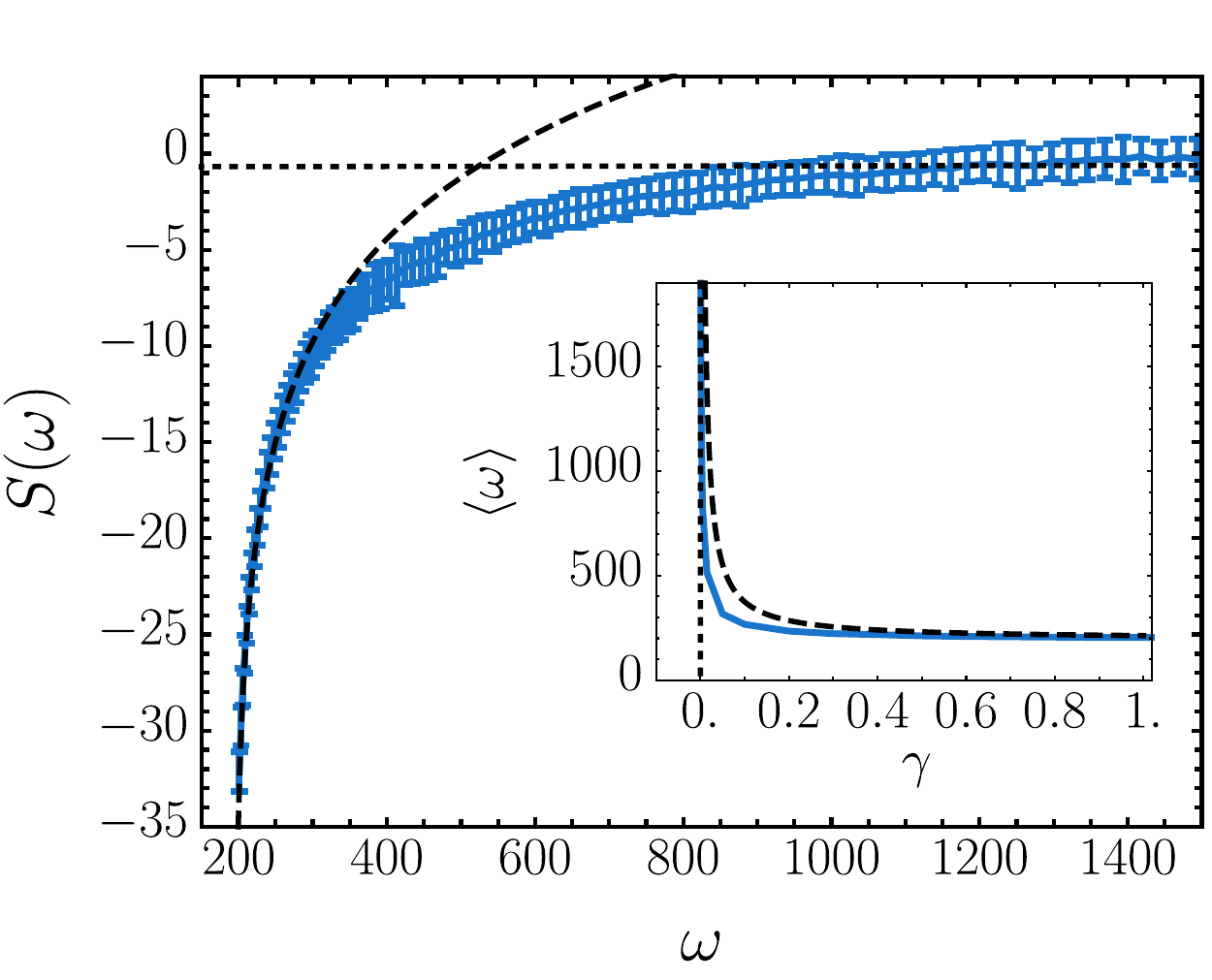}
\caption{\emph{Protocol entropy quantifies the diversity of protocols with average dissipation $\omega$.}
By sampling the distribution~\eqref{eq:protocolbias} with $t_{\rm obs} = 200$ Monte Carlo sweeps and using various biasing strengths $\gamma$, the protocol entropy $S(\omega)$ was computed using the Multistate Bennett Acceptance Ratio method~\cite{Shirts2008}.
The slope of the protocol entropy at $\omega = \left<\omega\right>$ gives the strength of the bias necessary to yield mean dissipation $\left<\omega\right>$. 
Asymptotic limits for small and large gamma are plotted as the dotted and dashed lines, respectively.
Inset: The average dissipation for protocols as a function of the bias $\gamma$.
Note that a small bias produces most of the achievable dissipation reduction.
}
\label{fig:entropy}
\end{figure}

The protocol entropy determined by sampling magnetization inversion dynamics is shown in Fig.~\ref{fig:entropy}.
Near the apparent minimum dissipation, a small increase in mean dissipation is accompanied by a steep rise in $S(\omega)$, i.e., the number of protocols grows rapidly as we permit modest excess dissipation.
In fact, the density of protocols increases by several orders of magnitude for a change in dissipation that is very small relative to dissipation fluctuations for a fixed protocol.
This rapid initial growth is captured well by the asymptotic form of Eq.~\ref{eq:asymp} which depends on the number of degrees of freedom in the protocol. 
Farther from the minimal value of mean dissipation, the protocol entropy climbs much more gradually.

The slope of $S(\omega)$ reflects the strength of bias $\gamma$ needed to depress the average dissipation.
The inset of Fig.~\ref{fig:entropy} illustrates a crossover between two regimes: small biases greatly reduce the mean dissipation but further reduction requires very large biases.  
Thus, weak biases on $\avg{\omega}_\Lambda$ can be greatly effective at directing the protocol sampling toward the optimum.  
We anticipate that these limiting behaviors are generic, and the corresponding asymptotic forms are derived in the SI.
The reduction in dissipation due to small values of $\gamma$ is governed by the variability of $\langle \omega \rangle_\Lambda$ in an unbiased protocol ensemble. 
Because complex systems typically depend sensitively on one or more of their control parameters, this variability should be substantial in general. 
Large values of $\gamma$ favor nearly optimal protocols, whose diversity is well described by~\eqref{eq:asymp}. 
Correspondingly, the mean dissipation in the large $\gamma$ limit decays slowly as $\langle \omega \rangle = \omega^* + n/(2\gamma).$

\subsubsection*{Gaussian fluctuations}
\begin{figure}[th]
\centering
\includegraphics[width=0.85\linewidth]{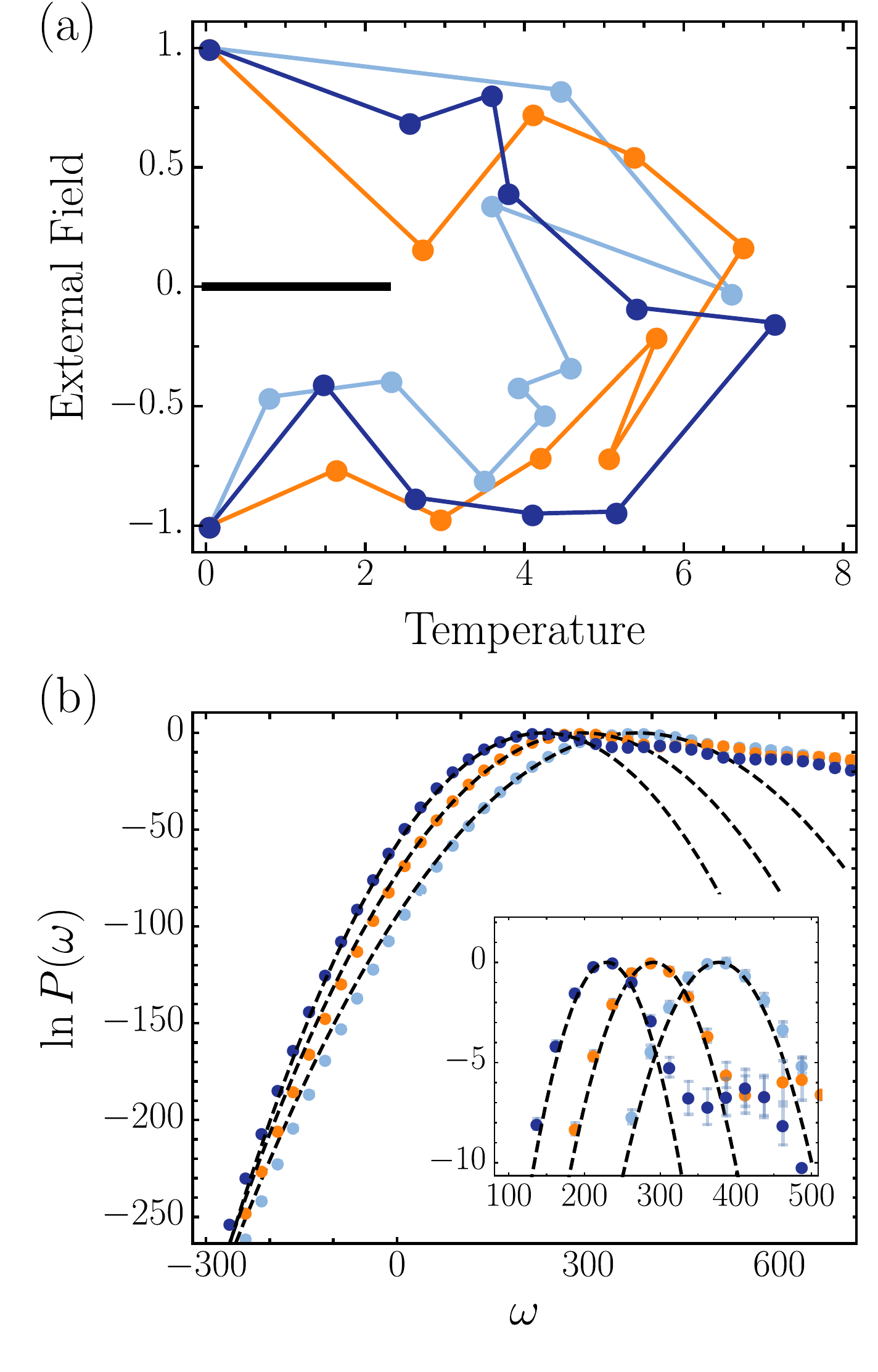}
\caption{\emph{Dissipation distributions for representative protocols.}
(a) Three randomly selected protocols from the ensemble~\eqref{eq:Nprotocolbias} are plotted on the temperature, external field plane as in Fig.~\ref{fig:protocols}.
(b) The distribution of dissipation values $P(\omega | \Lambda)$ for the three protocols, displayed with corresponding colors. 
Dashed black lines show Gaussian distributions with the same means $\left<\omega\right>_\Lambda$ as the sampled distributions and with variances $2 \left<\omega\right>_\Lambda$.
Inset: The neighborhood around the average dissipation values is shown in greater detail.}
\label{fig:diss}
\end{figure}

We have computed $S(\omega)$ for the spin inversion process both with and without the simplifying assumption that the dissipation is Gaussian distributed.
We find that the Gaussian approximation provides an estimate that is accurate within statistical error despite requiring a significantly reduced number of trajectories.
To more explicitly demonstrate the validity of the approximation, we selected three protocols from our sampling, which are shown in Fig.~\ref{fig:diss}(a). 

For each protocol, we computed the dissipation distribution $P(\omega|\Lambda)$.
Empirically, we find that these distributions are strikingly Gaussian over a large range of $\omega$ that includes~$\omega=0$.
At large positive values of dissipation, we observe ``fat'' exponential tails, consistent with the structure of generic current large deviation functions that has recently been demonstrated for the case of time-independent driving~\cite{Gingrich2016,Pietzonka2016}.
This fat tail, associated with clusters of spins that resist reorientation, only weakly restricts our use of the Gaussian dissipation assumption.
The positive $\lambda$ bias, useful to study low-dissipation behavior, focuses the sampling towards the Gaussian region of the distribution, rendering the exponential tails insignificant.

\section*{Discussion}
Low dissipation, the focus of our exploration of driving protocols, is one of many possible objectives for nonequilibrium control.  
Indeed, minimizing dissipation can be viewed as an instance of the extensively studied problem of stochastic optimal control~\cite{Fleming1971,Fleming1977,Dupuis2004,Hartmann2012, Chetrite2015}.
The formulation of stochastic optimal control as a path integral control problem~\cite{Kappen2005prl,Theodorou2010,Chernyak2014,Thijssen2015}  has a particularly close connection to our work, in that importance sampling of trajectories can be used to iteratively refine a protocol towards the optimum~\cite{Zhang2014,Kappen2016}.
A distinct feature of our approach is the systematic \emph{sampling} of protocols, as opposed to the more limited goal of strictly \emph{optimizing} them. 
By simultaneously sampling trajectories and protocols with a well-defined bias, we identify low-dissipation protocols without the challenging task of converging trajectory-space averages for any particular protocol, as is required in iterative optimization methods. 
Protocols harvested by our procedure, in contrast to those encountered in an optimization, quantitatively reflect the diversity of low-dissipation protocols.

Scrutinizing near-optimal protocols complements the search for optimality in several ways.  
In simple model systems that can be optimized exactly, minimum-dissipation protocols are known to involve features that are singular or may be impractical to implement~\cite{Schmiedl2007}.  
In such cases, the collection of near-optimal protocols becomes a natural target for design.  
Even when the optimal protocol may be physically achieved, its form does not directly indicate which features of a driving history are essential to its success, and which are irrelevant. 
In our approach, relevant features can be readily identified through their limited variability in the protocol ensemble, as illustrated by the cloud of protocols in Fig.~\ref{fig:protocols}(b).

Finally, we note that efficient but sub-optimal nonequilibrium transformations are almost certainly the norm in biology at many scales.
Indeed, the evolutionary dynamics of biological adaptation might be viewed as an importance sampling on the space of protocols, roughly akin to the sampling methods developed in this paper. 
The surprising, often eccentric strategies used to perform simple tasks in biology are, perhaps, indicative of the myriad options provided by an ensemble of protocols evolving under a complex set of constraints.

\matmethods{
We sample the joint space of trajectories and protocols using Markov Chain Monte Carlo methods.  
Each point in this space consists of a protocol and $N$ independent trajectories subject to that protocol.  
With tunable bias $\lambda$, the Markov chain samples the distribution in~\eqref{eq:Nprotocolbias}.  
We restrict the the space of protocols with a view towards experimental practicality.
For the 2D Ising example, we impose two such restrictions.  
Firstly, to allow only slowly varying protocols, we parameterize our protocol space by the values of temperature and external field at 11 evenly spaced times.  
We call the values at these special times the control points.  
Between any two neighboring control points the temperature and field strength are linearly interpolated.  
Secondly, we require that $-1\leq h \leq 1$ and $0 \leq T \leq 8$ for all control points to mimic that physical apparatuses can tune controls over bounded ranges.
Limiting the protocol space in these ways can be viewed as a regularization scheme that simplifies the representation problem of optimal control theory~\cite{Kappen2016}.  
We note, however, that the protocol entropy depends on our choice of regularization.
If, for example, we were to use many more control points with the same $t_{\rm obs}$, then we would introduce many additional protocols with high-frequency features.
Consequently, $S(\omega)$ would grow more rapidly, following Eq.~\ref{eq:asymp}.

Each Monte Carlo move first attempts to adjust the protocol by moving a single control point by a random displacement in the temperature-field plane.  
The move is constructed to be symmetric, meaning the probability of selecting any displacement vector equals the probability of a displacement of the opposite sign.  
Using this trial protocol, $N$ new trajectories are simulated using a sequence of Glauber single-spin flips.  
Conventionally, each step of the Glauber Ising dynamics chooses a random spin, which is flipped to generate a trial configuration.  
To enable more efficient noise-guided trajectory sampling, we use a modified Glauber dynamics: the trial configuration is given by setting the randomly selected spin to either the up or the down state without reference to its prior state~\cite{Gingrich2015}.
The move is futile when the selected spin is already in the trial configuration.  
Since half of the moves are futile on average, the Monte Carlo time is rescaled by a factor of two as compared to ordinary single-spin-flip Glauber dynamics.  
Following each move, the probability of running that step backwards is computed, enabling an explicit calculation of the dissipation of each trajectory.  
The new protocol and trajectories are conditionally accepted with probability $\text{min}[1, \exp(-\lambda \Delta \overline{\omega})]$, where $\Delta \overline{\omega}$ is the difference between the sample means under the original and trial protocols.

The protocol entropy is calculated, up to a constant offset $\ln \Omega_0$, using a weighted average over the protocols collected by the Monte Carlo procedure,
\begin{equation}
S(\omega) = \ln \left[\Omega_0 \int d\Lambda(t) \ \frac{P_{\rm sampled}[\Lambda(t)]}{\left<e^{-\lambda \omega / N}\right>_\Lambda^N} \delta\left(\omega - \left<\omega\right>_\Lambda\right)\right].
\label{eq:protocolentropy2}
\end{equation}
From a set of $M$ sampled protocols, $\left\{\Lambda_1, \Lambda_2, \hdots, \Lambda_\alpha, \hdots, \Lambda_M\right\}$, we therefore estimate
\begin{equation}
S(\omega) = \ln \left[ \frac{\Omega_0}{M} \sum_{\alpha=1}^M \frac{\delta\left(\omega - \left<\omega\right>_{\Lambda_\alpha}\right)}{\left<e^{-\lambda \omega / N}\right>_{\Lambda_\alpha}^N} \right].
\label{eq:hist}
\end{equation}
Operationally, this amounts to collecting a histogram of values of $\left<\omega\right>_\Lambda$ with each entry weighted by the corresponding value of $\left<e^{-\lambda \omega / N}\right>_\Lambda^N$.
To generate Fig.~\ref{fig:entropy}, each of these weights is computed by estimating the exponential average from 1000 independent trajectories.
In practice, $S(\omega)$ is constructed using the Multistate Bennett Acceptance Ratio (MBAR) method to combine samples collected with $N = 20$ and with several different values of the bias ranging from $\lambda = 0$ to $\lambda = 1$.
The offset $\ln \Omega_0$ is chosen such that $S(\omega)$ is zero at its maximum.

The protocol entropy can be computed much more efficiently when the Gaussian approximation may be used to evaluate the exponential average.
To evaluate the validity of this approximation for the Ising dynamics, we compute the actual dissipation distributions by sampling trajectories with a fixed protocol.
These trajectories are importance sampled using harmonic biases, which restrain the dissipation to fluctuate around a specified value.
By choosing several different harmonic biases, trajectories were biased into both tails of the dissipation distribution, which was reconstructing using the MBAR method~\cite{Shirts2008}.

When the Gaussian approximation is appropriate, it is wasteful to use a large value of $N$.
Low-dissipation protocols may be sampled with an effective biasing strength $\gamma = \lambda(1 - \lambda/N)$ using various combinations of $N$ and $\lambda$, and a small $N$ reduces the computational expense.
However, when $N$ is too small or $\lambda$ too large, the Monte Carlo acceptance probability drops precipitously, a fact elaborated upon in the SI.
Sampling efficiency is poor under these conditions because the Markov chain favors a collection of rare trajectories with dissipation below the mean (and often below zero).
This issue can be partially alleviated by introducing replica exchange moves\textemdash random swaps exchanging replicas with different biasing strengths $\lambda$.
The implementation of this procedure naturally mirrors the use of replica exchange to surmount kinetic traps when sampling low-temperature molecular configurations.
Further performance enhancements are obtained when trial trajectories are generated with random numbers (noises) which correlate with the noises of the previous collection of trajectories.
An implementation of this noise-guided sampling is described in detail elsewhere~\cite{Gingrich2015, Gingrich2015thesis}.
The noise guidance technique is not strictly required to perform the protocol sampling, but in the SI we show that it can provide significant practical benefits.
}
\showmatmethods 

\acknow{TRG acknowledges support from the NSF Graduate Research Fellowship, the Fannie and John Hertz Foundation, and the Gordon and Betty Moore Foundation as an MIT Physics of Living Systems Fellow.
GMR would like to acknowledge support from the NSF Graduate Research Fellowship. 
PLG was supported by the U.S. Department of Energy, Office of Basic Energy Sciences, through the Chemical Sciences Division (CSD) of the Lawrence Berkeley National Laboratory (LBNL), under Contract DE-AC02-05CH11231. 
GEC acknowledges support from the U.S. Army Research Laboratory and the U.S. Army Research Office under Contract No. W911NF-13-1-0390.}

\showacknow 


\bibliography{refs}

\newpage
\beginsupplement
\section*{Supporting Information (SI)}

\subsection{General features of the protocol entropy}
In the limit of large values of the bias parameter $\gamma,$ the difference between a sampled protocol $\Lambda(t)$ and the optimal protocol $\Lambda^*(t)$ should be small. 
If the average dissipation $\avg{\omega}_\Lambda$ is a smooth functional of the protocol, we can approximate the deviation from the minimum average dissipation in terms of the small protocol variations $\delta \Lambda = \Lambda - \Lambda^*$,
\begin{equation}
\avg{\omega}_\Lambda = \omega^* + \frac{1}{2} \sum_{i,j=1} \delta\Lambda_i\cdot  \mathbf{K}_{ij} \cdot\delta\Lambda_j,
\end{equation}
where the indices $i$ and $j$ label times at which the protocol is manipulated.
In this limit, we can explicitly calculate the moment generating function for the average dissipation, 
\begin{equation}
Z(\gamma) = \left\langle e^{-\gamma \avg{\omega}_\Lambda} \right\rangle \underset{\gamma \to \infty}{\sim} \frac{1}{\sqrt{\det(\gamma \mathbf{K})}} e^{-\gamma \omega^*} \propto \gamma^{-n/2} e^{-\gamma \omega^*}.
\end{equation}
and $n$ denotes the total number of degrees of freedom in the protocol.

\begin{figure}[t]
\centering
\includegraphics[width=0.9\linewidth]{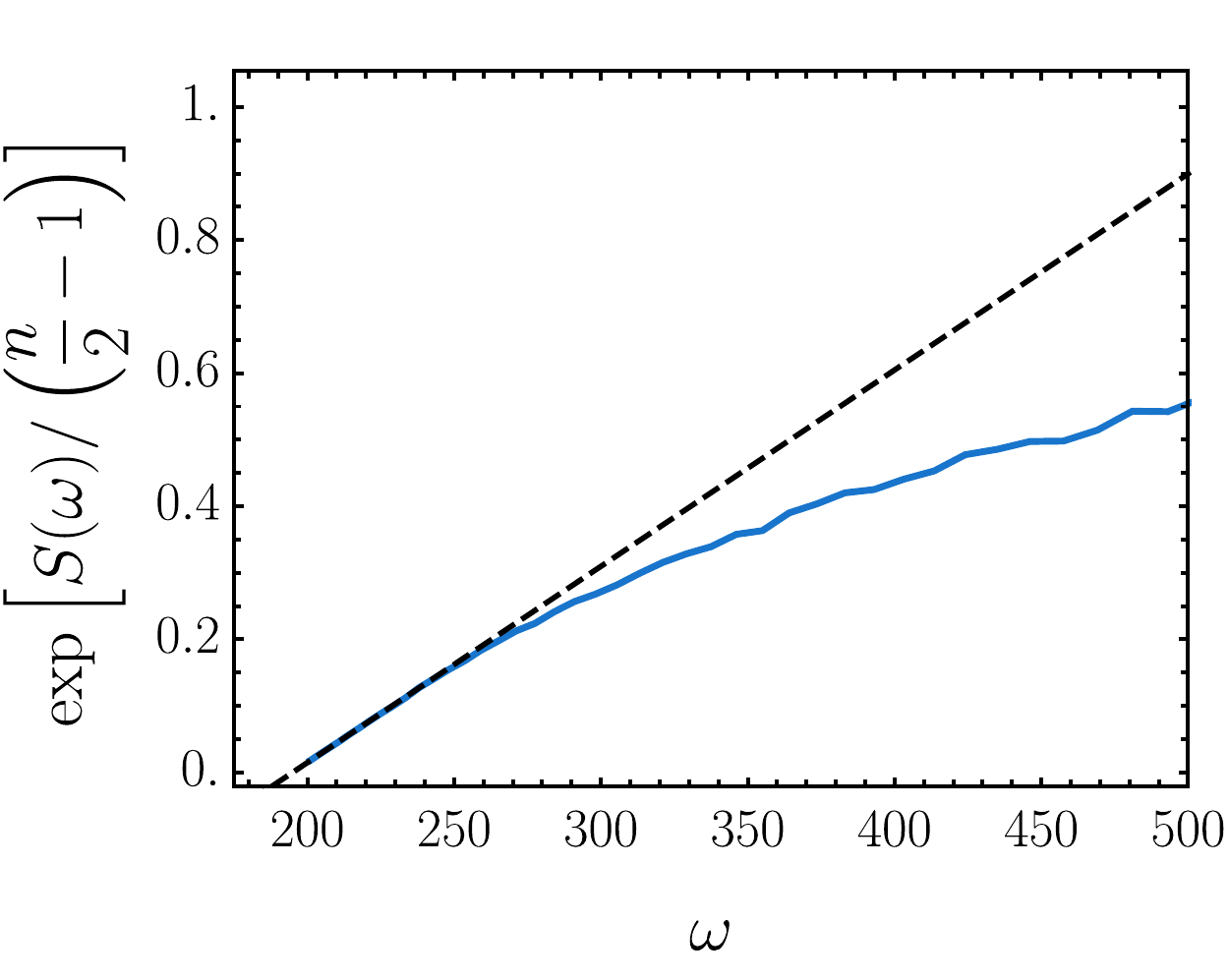}
\caption{
\emph{Asymptotic analysis of the Ising inversion protocol entropy.} 
The protocol entropy $S(\omega)$ from Fig.~\ref{fig:entropy} is transformed and plotted in blue to highlight the large $\gamma$-small $\omega$ limiting behavior. 
As described in Methods and Materials, the protocol entropy calculations used nine control points, each with two degrees of freedom.
The dashed black line is the asymptotic form given in~\eqref{eq:SIasymp} with $n = 18$.
}
\label{fig:entropy_asymp}
\end{figure}

The unbiased distribution of average dissipations $P(\omega) \propto \exp\bigl[ S(\omega) \bigr]$ is an inverse Laplace transform of $Z(\gamma),$
\begin{equation}
P(\omega) \propto (\omega - \omega^*)^{n/2 -1},
\end{equation}
and hence the entropy in the large $\gamma$ limit is, 
\begin{equation}
S(\omega) = \text{const} + \left(\frac{n}{2}-1\right)\ln( \omega - \omega^*).
\label{eq:SIasymp}
\end{equation}
Fig.~\ref{fig:entropy_asymp} compares the asymptotic expression~\eqref{eq:SIasymp} with the entropy computed for the Ising system.

From the asymptotic expression for $Z(\gamma)$, we can also compute the average dissipation associated with protocols sampled under the bias $\gamma$:
\begin{align}
\avg{\omega} &= -\d{}{\gamma} \ln Z(\gamma)\\
&= \omega^* + \frac{n}{2\gamma}.
\end{align}

In the absence of protocol bias, $\gamma=0$, typical values of dissipation are quite large. 
Provided that $\Lambda(t)$ is bounded, the mean dissipation $\langle \omega\rangle_0$ at $\gamma=0$ should nonetheless be finite, as is the corresponding variance $\langle(\omega - \langle \omega\rangle_0)^2\rangle_0$.
Sufficiently close to the maximum of $P(\omega)$ we therefore have
\begin{equation}
S(\omega) = S(\langle \omega\rangle_0) - 
{(\omega - \langle \omega\rangle_0)^2 \over
2 \langle
(\omega - \langle \omega\rangle_0)^2\rangle_0
}
\end{equation}
The parameters $\langle \omega\rangle_0$ and $\langle (\omega - \langle \omega\rangle_0)^2\rangle_0$ in this expression are determined numerically by computing average dissipation for protocols generated in the $\gamma = 0$ ensemble.
The corresponding curve is plotted in Fig.~\ref{fig:entropy} as a dotted line.

The corresponding moment generating function is 
\begin{equation}
Z(\gamma) \propto \exp\left[
-\gamma \langle \omega\rangle_0
+ {1\over 2}\langle (\omega -
\langle \omega\rangle_0)^2\rangle_0 \gamma^2
\right],
\end{equation}
giving a $\gamma$-biased average dissipation:
\begin{equation}
\langle \omega \rangle = 
\langle \omega\rangle_0 - \gamma \langle (\omega -
\langle \omega\rangle_0)^2\rangle_0,
\end{equation}
which is shown as a dotted line in the inset of Fig.~\ref{fig:entropy}.

\subsection*{Sampling efficiency}
Ergodic sampling of $P_\text{canon}[\Lambda(t)]$ requires that decorrelated protocols be generated, and the efficiency of the sampling depends on how many Monte Carlo moves are required to produce each decorrelated sample.
As discussed in the main text, the canonical distribution with bias $\gamma$ may be accessed with various choices of $N$ and $\lambda$, yet the sampling efficiency depends on these choices.
A large value of $N$ bears a clear computational cost since $N$ new trajectories must be simulated for each trial protocol.
It is not, however, always optimal to choose the minimal $N$ capable of generating bias $\gamma$, $N^* = \text{Ceil}(4\gamma)$.
The problem with using a small value of $N$ is that rare low-dissipation trajectories (including those with negative dissipation) can depress the probability of accepting a protocol move, even when the trial protocol has a lower \emph{average} dissipation.
The smaller the choice of $N$, the more rare trajectories influence the acceptance of protocol moves.
Consequently, the optimal choice of $N$ often exceeds $N^*$.

In practice, the best choice of $N$ depends on the Monte Carlo moves used to update trajectories and protocols.
For example, there is a trade-off in choosing the best protocol space moves.
Large changes to the protocol have a low acceptance probability, but small steps require many moves before sampling a protocol with a decorrelated value of mean dissipation.
To quantify the trade-offs between possible choices of $N$ and of Monte Carlo moves, we construct a correlation function:
\begin{equation}
C_\omega(j) = \left<\delta\left<\omega\right>_{\Lambda_i} \delta\left<\omega\right>_{\Lambda_{i+j}}\right>,
\label{eq:correlationfunction}
\end{equation}
where $\delta\left<\omega\right>_{\Lambda_i}$ is the difference between the $i^{\rm th}$ sampled protocol's average dissipation and the mean dissipation, found by averaging over all protocols in the ensemble.
Assuming exponentially decaying correlations ($C_\omega(j)/C_\omega(0) = e^{-j / \tau_\omega}$), we set the correlation time to the decay constant $\tau_\omega$.
This decorrelation time is the typical number of protocols which must be sampled before generating a protocol with a decorrelated value of the mean dissipation.
Since each new protocol must be accompanied by the simulation of $N$ new trajectories, the computational cost for obtaining each decorrelated sample is given by $N\tau_\omega$.

The decorrelation time $\tau_\omega$ implicitly depends on $N$ as well as on the details of the sampling moves.
To gain intuition about the most computationally-efficient choice of $N$, we first consider a simplified Gaussian model.
For this model, the optimal $N$ slightly exceeds $N^*$.
The Ising dynamics is more complicated, requiring $\tau_\omega$ to be computed from simulations.
We show that the computational expense depends on $N$ in a similar manner as in the Gaussian model, though the computational expense may be reduced by employing noise-guided path sampling.

\subsubsection*{Gaussian model}
We sample a scalar protocol $\Lambda$ in the vicinity of the minimum dissipation protocol $\Lambda^*$.
The advantage of sampling near $\Lambda^*$ is that we can Taylor expand to arrive at an expression for the average dissipation associated with a perturbation $\delta \Lambda = \Lambda - \Lambda^*$,
\begin{equation}
\left<\omega\right>_{\Lambda} = \left<\omega\right>_{\Lambda^*} + \frac{1}{2} K\dL^2.
\end{equation}
We assume that the dissipation distribution associated with a protocol $\Lambda$ is Gaussian,
\begin{equation}
p(\omega| \Lambda) \propto \exp \left(-\frac{\left(\omega - \left<\omega\right>_{\Lambda}\right)^2}{4{\left<\omega\right>_\Lambda}}\right).
\label{eq:gaussiandissipaitontoy}
\end{equation}
We furthermore use a Gaussian distribution to propose a trial protocol,
\begin{equation}
P( \Lambda_i \to \Lambda_{i+1}) \propto \exp\left(-\frac{(\Lambda_i - \Lambda_{i+1})^2}{2 D^2}\right).
\end{equation}
Using the new trial protocol, the dissipation for $N$ independent trajectories is drawn from~\eqref{eq:gaussiandissipaitontoy} to compute the sample mean dissipation for the new protocol $\overline{\omega}$.
For a choice of bias $\gamma$, we choose $\lambda$ and $N$ such that $\gamma = \lambda(1 - \lambda/N)$ and accept the new protocol and trajectories with probability
\begin{equation}
P_{\rm accept}^{\Lambda \to \Lambda'} = \min\left[1, e^{-\lambda \Delta \overline{\omega}}\right],
\label{eq:paccepttoy}
\end{equation}
where $\Delta \overline{\omega}$ is the difference between the new protocol's sample mean dissipation and that of the old protocol and trajectories.

\begin{figure*}[p]
\centering
\includegraphics[width=0.85\linewidth]{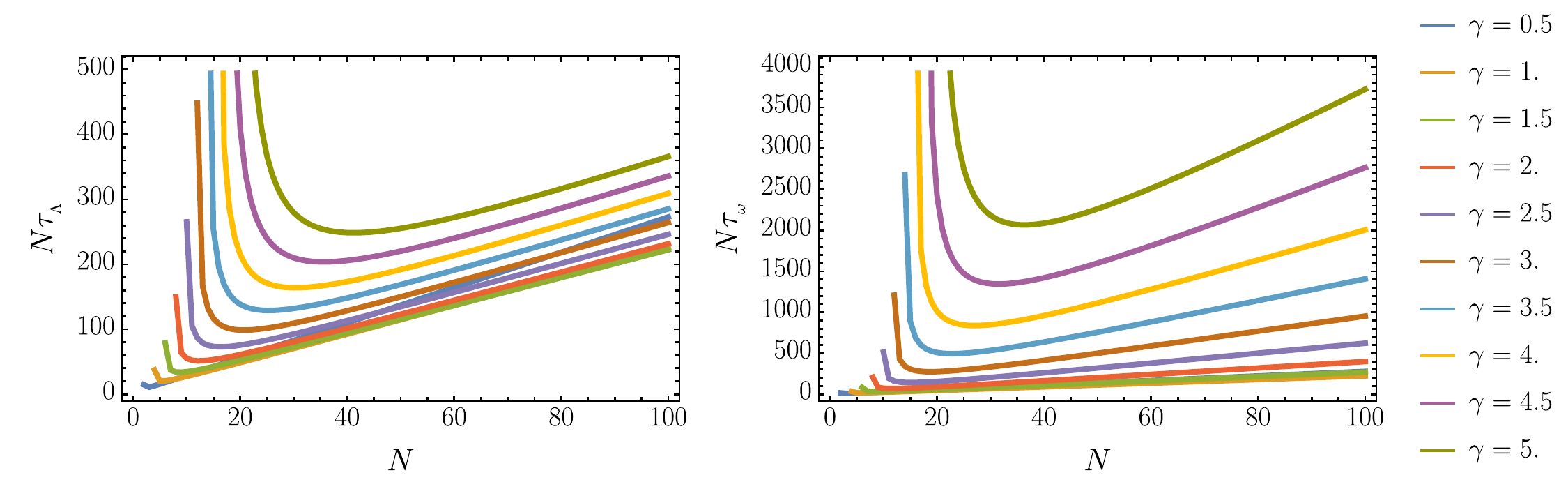}
\caption{\emph{Computational expense of sampling varies non-monotonically with $N$.}(Left) Computational cost, in number of trajectories needed to generate a decorrelated protocol, for the Gaussian model with various values of the bias $\gamma$. (Right) Cost to generate protocols with a decorrelated value of the mean dissipation.}
\label{fig:asymp}
\end{figure*}

The correlation between subsequent samples is given by
\begin{align}
\nonumber C_\omega(1) &= \int d\Lambda_i\ \int d\Lambda_{i+1} P(\Lambda_i) P(\Lambda_{i}\to \Lambda_{i+1}) \delta \left<\omega\right>_{\Lambda_i} \\
& \ \ \times \left( P_{\rm accept}^{\Lambda_i \to \Lambda_{i+1}} \delta\left<\omega\right>_{\Lambda_{i+1}} + \left(1-P_{\rm accept}^{\Lambda_i \to \Lambda_{i+1}}\right) \delta \left<\omega\right>_{\Lambda_{i}}\right).
\end{align}
with
\begin{align}
\nonumber
P_{\rm accept}^{\Lambda_i \to \Lambda_{i+1}} =& \int_{-\infty}^0 d\Delta\overline{\omega} \ P(\Delta \overline{\omega}|\Lambda_i, \Lambda_{i+1})\\
& \ \  + \int_{0}^\infty d\Delta\overline{\omega} \ P(\Delta \overline{\omega}|\Lambda_i, \Lambda_{i+1})e^{-\lambda \Delta \overline{\omega}}
\end{align}
giving the acceptance probability for the Monte Carlo protocol move.
Since the sample mean dissipations (for both the old and trial protocol) are Gaussian distributed, $P(\Delta \overline{\omega}|\Lambda_i, \Lambda_{i+1})$ is also a Gaussian,
\begin{align}
\nonumber P(\Delta\overline{\omega}|&\Lambda_i, \Lambda_{i+1}) \propto\\
& \exp\left(-\frac{N\left(\Delta\overline{\omega} - \left(\avg{\omega}_{\Lambda_{i+1}} - \left(1 - \frac{2\lambda}{N}\right)\avg{\omega}_{\Lambda_i}\right)\right)^2}{4\left(\avg{\omega}_{\Lambda_i} + \avg{\omega}_{\Lambda_{i+1}}\right)}\right).
\end{align}
The acceptance rate may therefore be expressed in terms of error functions.
The number of attempted protocol moves required to sample protocols with decorrelated mean dissipations is given by $\tau_\omega = -1/\ln (C_\omega(1)/C_\omega(0))$.
A complicated integral expression for $\tau_\omega$ in terms of $\left<\omega\right>_{\Lambda^*}, D, \gamma,$ and $N$ may be derived in terms of error functions.
Numerically evaluating the integral expression yields the $N$-dependence of the computational cost, $N\tau_\omega$, for generating protocols with decorrelated average dissipation.
Analogously, we may compute the time to find decorrelated values of the scalar protocol $\tau_\Lambda$ as the decay constant for the correlation function
\begin{equation}
C_\Lambda(j) = \left<\delta \Lambda_i \delta \Lambda_{i+j}\right>.
\end{equation}
Both $\tau_\omega$ and $\tau_\Lambda$ yield the same heuristic: the computational expense is minimal when $N$ slightly exceeds $N^*$, as plotted in Fig.~\ref{fig:asymp}.

\subsubsection*{Ising dynamics}
Using various choices of $\lambda$ and $N$, the canonical protocol ensemble is sampled by a sequence of protocol moves which alter the control point at a single time.
Either the control point's magnetic field strength is increased by a random uniform number between $-0.5$ and $0.5$ or the temperature is increased by a random uniform number between $-2.5$ and $2.5$.
The protocol move is resampled if the magnitude of the magnetic field strength at any time exceeds 1 or if the dimensionless temperature exits the range $[0, 8]$.
$N$ trajectories are generated by repropagating forward dynamics from the initial time or by running time-reversed dynamics from the final time slice and reweighting the trajectories into the forward-trajectory ensemble.
When noise guidance path sampling is not being used, each trial trajectory is generated by drawing new random noises to carry out the steps of the spin-flip dynamics.
The noise guidance path sampling scheme recycles, with probability $1-\epsilon$, each random number from the previous trajectory.
Over time, each random number gets resampled uniformly, but the correlations between an old trajectory and a new trajectory are enhanced~\cite{Gingrich2015, Gingrich2015thesis}.

\begin{figure*}[p]
\centering
\includegraphics[width=0.8\linewidth]{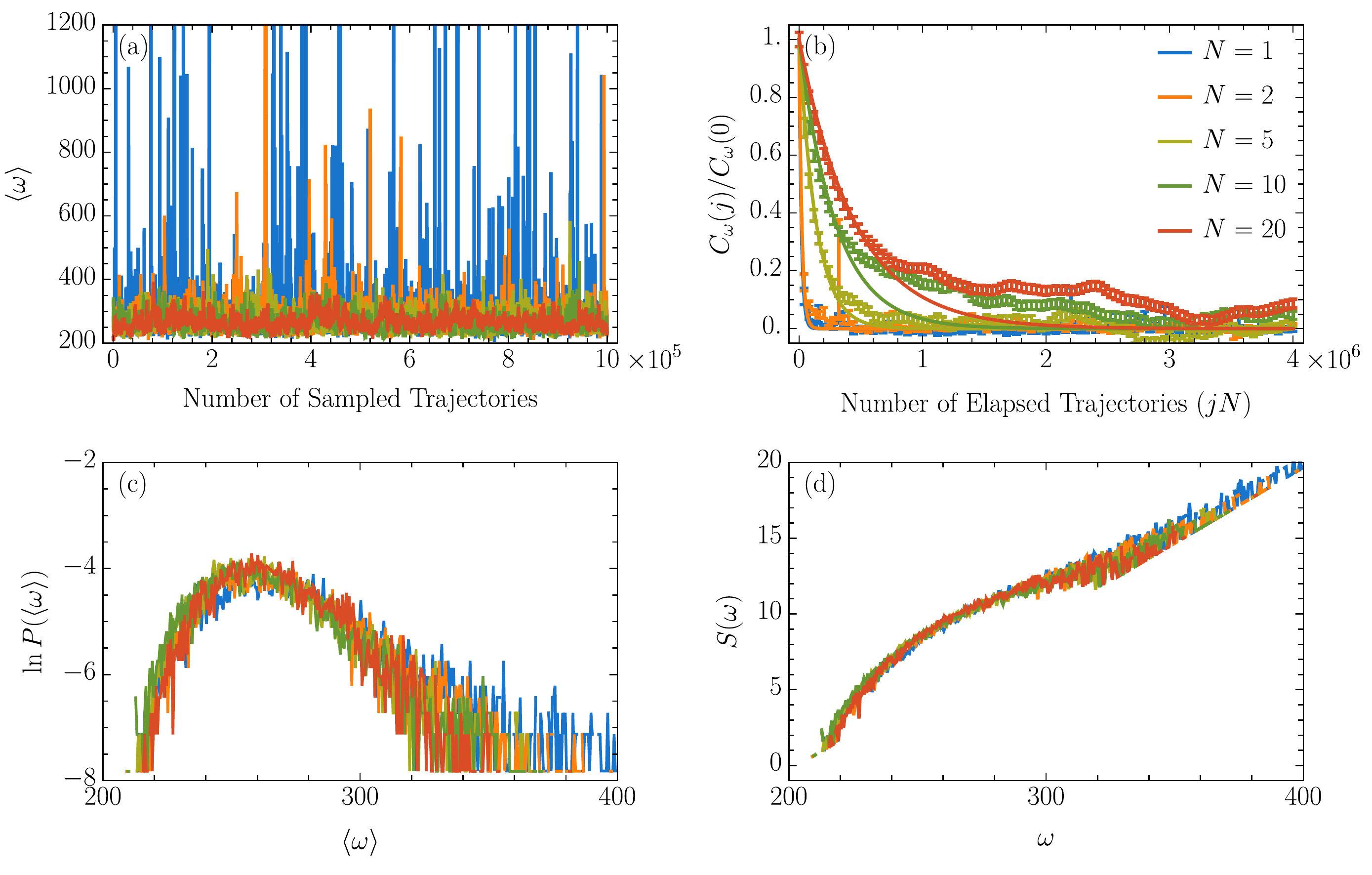}
\caption{
\emph{Sampling $e^{-\gamma \left<\omega\right>_\Lambda}$ with $\gamma = 0.1$ and various choices of $N$.}
To analyze the marginal protocol distributions generated from different choices of $N$ (these marginal distributions agree with each other in the Gaussian dissipation distribution limit), we report on the statistics of $\left<\omega\right>_\Lambda$, estimated as a sample mean of 1000 independent trajectories.
(a) Fluctuations in this mean dissipation for different sampled protocols.  
The $x$-axis measures ``time'' for the protocol sampling Monte Carlo procedure by counting the total number of simulated spin-flip trajectories,which differs from the total number of sampled protocols by a factor of $N$.
Counting the number of trajectories roughly reflects the computational expense.
(b) Correlation functions showing how many trajectories must be simulated before a new protocol is produced with a statistically uncorrelated value of the mean dissipation.
Using large values of $N$ can be unnecessarily wasteful because each protocol requires sampling $N$ trajectories.
(c) Histogram of the sampled average dissipation values.
Note that the different choices of $N$ produce the same distributions except that extreme events can be over-represented by small-$N$ sampling.
(d) The histogram of (c) is reweighted to yield the protocol entropy.
In the neighborhood of the histogram's peak the protocol entropy calculation is robust to errors from small-$N$ sampling.
By choosing various biasing strengths $\gamma$,  we highlight different ranges of $\left<\omega\right>$ and stitch them together to compute the protocol entropy curve reported in the main text.
}
\label{fig:samplingoverview}
\end{figure*}

\begin{figure*}[p]
\centering
\includegraphics[width=0.65\linewidth]{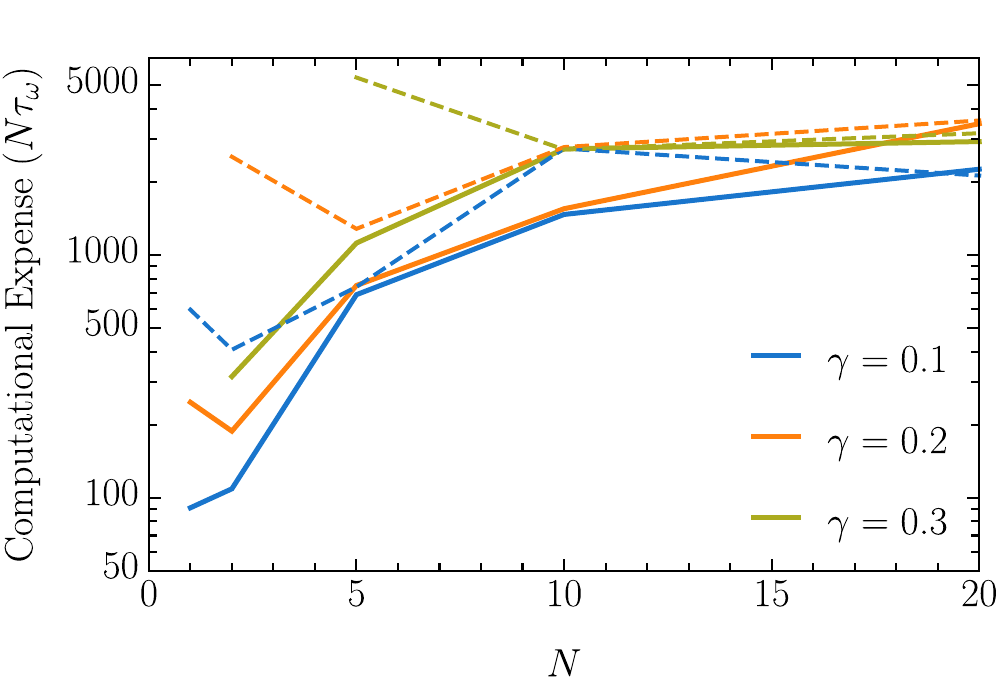}
\caption{
\emph{Computational expense of generating decorrelated protocols.}
For various $N$ and $\gamma$, an exponential time constant $\tau$ was extracted from the correlation functions plotted in Fig.~\ref{fig:samplingoverview}(b) to obtain the number of trajectories which must be simulated to obtain a decorrelated protocol.
This computational expense depends not only on $N$ and $\gamma$ but also on the manner in which new trajectories are generated.
The dashed lines show the expense when each trajectory is drawn at random.
Solid lines show the improvement that can be achieved by using noise guidance.
When generating a new Ising trajectory, the noise guidance scheme re-used each of the previous trajectory's random numbers with probability 0.999 ($\epsilon = 10^{-3})$.
}
\label{fig:computationalexpense}
\end{figure*}

To assess how rapidly the protocol space is sampled, protocols were stored and an accurate estimate of $\left<\omega\right>_\Lambda$ was found by averaging over 1000 trajectories with the fixed protocol.
Fig.~\ref{fig:samplingoverview}(a) shows the (Monte Carlo) time series of this average dissipation.
Using this time series, the correlation function $C(j)$ is computed for each choice of $N$ and $\lambda$.
The correlation functions are fit to exponential decays, and the decay constant $\tau$ is extracted to yield the computational cost $(N\tau)$ as a function of $N$.
Fig.~\ref{fig:computationalexpense} illustrates that the optimal choice of $N$ exceeds $N^*$, as in the Gaussian model.
The noise guidance strategy aids small-$N$ sampling, thereby lowering the optimal $N$.
For these small choices of $N$, the noise guidance scheme offers a roughly order of magnitude speed up.

\subsection*{Sample mean fluctuations}
The biasing strength $\gamma$ necessary to sample protocols with dissipation $\left<\omega\right>$ is given by the slope of $S(\omega)$ at $\omega = \left<\omega\right>$.
When $S(\omega)$ is especially steep, it therefore requires a very strong bias to access the low-dissipation entropy.
Sampling efficiency worsens for large $\gamma$, so we note an alternative method for computing $S(\omega)$ that makes use of the fluctuations in the sample mean of $N$ trajectories.

\begin{figure*}[p]
\centering
\includegraphics[width=0.65\linewidth]{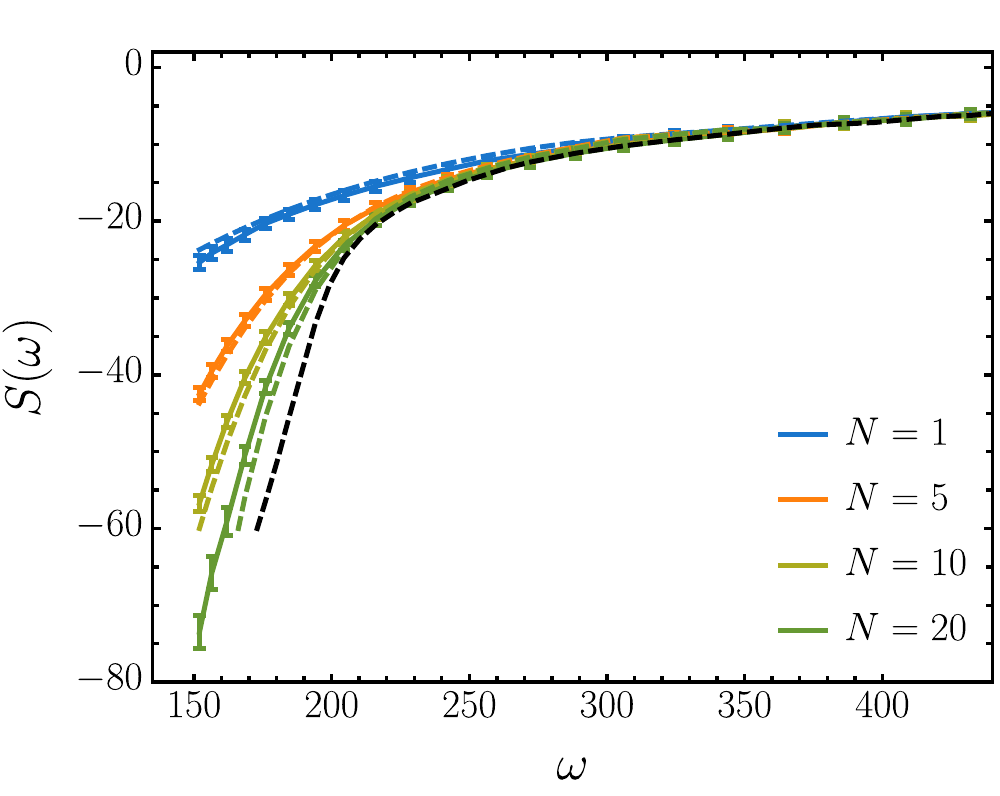}
\caption{
\emph{Inferring protocol entropy from finite $N$ sampling.}
The colored lines with error bars show $\overline{S}(\omega)$ for four choices of $N$, computed using MBAR to combine sampling with various choices of $\gamma$.
The form of $S(\omega)$ is inferred (black dashed line) by fitting to~\eqref{eq:infer} for $N =1, 5, 10, 20$.
The dashed colored lines show the expected $\overline{S}(\omega)$ given the inferred protocol entropy.
It is typically harder to sample in regions where $\overline{S}(\omega)$ is steep.
Since the entropy is less steep for small $N$, it can be productive to compute $\overline{S}(\omega)$ with modest $N$ then infer the $N\to \infty$ limit $S(\omega)$.
}
\label{fig:samplemeanfitting}
\end{figure*}

We define the entropy $\overline{S}(\omega)$ to give the density of protocols and trajectories which have a sample mean dissipation within an infinitesimal window of $\omega$:
\begin{equation}
\overline{S}(\omega) = \ln \left[\overline{\Omega}_0\int d\Lambda(t) dx_1(t) \hdots dx_N(t) \ \delta \left(\sum_i \frac{\omega\left[x_i(t), \Lambda\right]}{N} - \overline{\omega}_\Lambda\right)\right].
\end{equation}
In the limit $N \to \infty$ this entropy must converge to the protocol entropy, $S(\omega)$, but $\overline{S}(\omega)$ falls off less rapidly around the minimal average dissipation.
The tails of $\overline{S}(\omega)$ with low sample mean may therefore be importance sampled with a bias $\lambda.$ 
Sampling $S(\omega)$ directly, on the other hand, requires the stronger bias $\gamma$.
By making the Gaussian approximation, we can reconstruct $S(\omega)$ from $N-$dependence of $\overline{S}(\omega)$ using
\begin{equation}
\overline{S}(\omega) = \ln \left[\overline{\Omega}_0 \int d\omega \ \exp\left(-\frac{N(\overline{\omega}-\omega)^2}{4 \omega} + S(\omega)\right)\right],
\label{eq:infer}
\end{equation}
a technique illustrated in Fig.~\ref{fig:samplemeanfitting}.

\end{document}